\documentclass{aa}   
\usepackage{graphicx}
\usepackage{epstopdf}
\usepackage{soul}
\usepackage[varg]{txfonts}
\usepackage{color}
\usepackage{natbib}
\bibpunct{(}{)}{;}{a}{}{,}
\begin{document}

\title{Fossil groups in the Millennium simulation}
\subtitle{From the brightest to the faintest galaxies during the past 8 Gyr}
\author{Mar\'ia Jos\'e Kanagusuku\inst{1}\fnmsep\thanks{mjkanagusuku@gmail.com}  
\and Eugenia D\'iaz-Gim\'enez\inst{1,2} 
\and Ariel Zandivarez\inst{1,2}}

\institute{
Instituto de Astronom\'{\i}a Te\'orica y Experimental, IATE, CONICET, C\'ordoba, Argentina
\and
Observatorio Astron\'omico, Universidad Nacional de C\'ordoba, Laprida 854, X5000BGR, 
C\'ordoba, Argentina
}
\date{Received XXX; accepted XXX}
%%%%%%%%%%%%%%%%%%%%%%%%%%%%%%%%%%%%%%%%%%%%%%%%%%%%%%%%%%%%%%%%%%%%%%%%%%%%%%%
\abstract{}
{We investigate the evolution of bright and faint galaxies in fossil and non-fossil groups.}
{We used mock galaxies constructed based on the Millennium run simulation II.
We identified fossil groups at redshift zero according to two different 
selection criteria, and then built reliable control samples 
of non-fossil groups that
reproduce the fossil virial mass and assembly time distributions. 
The faint galaxies were defined as having r-band absolute magnitudes in the range [-16,-11].
We analysed the properties of the bright and faint galaxies in fossil 
and non-fossil groups during the past 8 Gyr. 
}
{
We observed that the brightest galaxy in fossil groups is typically brighter and
more massive than their counterparts in control groups. 
Fossil groups developed their large magnitude gap between the brightest galaxies 
around 3.5 Gyr ago. The brightest galaxy stellar masses 
of all groups show a notorious increment at that time. 
By analysing the behaviour of the magnitude gap between the first 
and the second, third, and fourth ranked galaxies, we found that at earlier times,
fossil groups comprised two large brightest galaxies 
with similar magnitudes surrounded by much fainter galaxies, while
in control groups these magnitude gaps were never as large as in fossils.
At early times, fossil groups in the faint population were denser
than non-fossil groups, then this trend reversed, and finally they became 
similar at the present day. 
The mean number of faint galaxies in non-fossil systems
increases in an almost constant rate towards later times, while this 
number in fossil groups reaches a plateau at $z\sim0.6$ that 
lasts $\sim 2$ Gyr, and then starts growing again more rapidly. 
}
{The formation of fossil groups is defined at the very beginning
of the groups according to their galaxy luminosity sampling, 
which could be determined by their merging rate at early times.}

%\titlerunning{Faint galaxy population inhabiting fossil groups}
\maketitle
%%%%%%%%%%%%%%%%%%%%%%%%%%%%%%%%%%%%%%%%%%%%%%%%%%%%%%%%%%%%%%%%%%%%%%%%%%%%%%%
\section{Introduction}   
The true nature of fossil groups in the Universe still puzzles the astronomical
community. These peculiar systems are one of the most intriguing places in the 
Universe where giant elliptical galaxies are hosted.  

Since their definition at the beginning of the past decade \citep{jones03},
the existence of these systems with a very luminous X-ray source 
($L_X \ > \ 10^{42} \ h^{-2}_{50} \ erg \ s^{-1}$)
and a very optically dominant central galaxy (magnitude gap between the two
brightest galaxies, $\Delta M_{12}$, greater than 2), 
many studies were performed to unveil their formation scenario.
Several of these attempts have intended to quantify their incidence rate,
dynamical masses, physical properties, etc. (see for instance, 
\citealt{mendes06,cypriano06,khosroshahi06a,khosroshahi06b}).
A special mention should be given to a recent effort to collect observational
evidence to study fossil systems, which it is known as the
``Fossil Group Origins'' project. This is a collaboration to study galaxy
systems previously identified as fossil groups by \cite{fogo0},
which has attempted to address several questions such as
studying high-redshift massive systems and their fossil-like behaviour \citep{fogo1}, 
the intrinsic difference between the brightest central galaxies in fossils and normal 
galaxy systems \citep{fogo2}, the correlation between their optical and X-ray luminosity
\citep{fogo3}, confirming the fossil nature of part of the original
group sample \citep{fogo4},
and analysing the dependence of the luminosity function
on the magnitude gap \citep{fogo5}.  

There is another approach to understand the real nature of these peculiar galaxy
systems, and that is through numerical experiments. From some of these studies
carried out using numerical simulations, we were able to deepen our 
understanding of the different formation scenarios (see for instance 
\citealt{donghia05,vonbenda08}). 
When these experiments are performed using a combination of a large cosmological 
simulation and a semi-analytical model of galaxy formation, very interesting analyses
can be done. In the past years, several studies have used synthetic 
galaxies to analyse the evolution of fossil groups. 
Particularly, very reliable results were obtained for those semi-analytical surveys
constructed based on one of the largest numerical simulations currently available,
the Millennium simulation (\citealt{springel05}, hereafter MS). Using this tool,
\cite{dariush07} were the first to confirm that fossil systems identified in the MS 
assembled a larger portion of their masses at higher redshifts than non-fossil groups,
suggesting that the most likely scenario for fossil groups is that they are not
a distinct class of objects, but simply examples of systems that collapsed earlier.  
In a later work, \cite{dariush10} suggested refinements to the fossil definition 
to enhance its efficiency in detecting old systems. 
On the other hand, in the first work of this series, \cite{diaz08}
studied the evolution of the first-ranked galaxies in the MS, 
finding that despite the earlier assembly time of fossil systems, 
first-ranked galaxies in fossil groups assembled half of
their final mass and experienced their last major merger
later than their non-fossil counterparts, implying that they followed a 
different evolutionary pathway. In a second work, \cite{diaz11}
intended to characterise the outskirts of fossil groups
in contrast with those observed in normal groups. They observed that the
environment was different for fossil and non-fossil systems with similar masses
and formation times along their evolution, encouraging the idea that their
surroundings could be responsible for the formation of their large magnitude gap.
Hence, consensus has clearly yet to be reached regarding the nature of fossil systems.

Some of the formation scenario proposed for fossil systems led us 
to feed a particular working hypothesis: 
that the population of faint galaxies inhabiting these
galaxy systems might have undergone a different evolutionary history than
is expected in normal systems. Several works have intended to understand the
role of the faint galaxy population in fossil groups. For instance, 
the early work of \cite{donghia04}, who suggested that it is 
expected that fossil groups may lack faint galaxies, 
in what they called the missing satellite problem in 
cold dark matter cosmologies. Further analysis performed using the luminosity function
of fossil group galaxy members by \cite{mendes09} has shown that there is no
significant evidence that this problem with faint galaxies actually occurs. In addition,
\cite{sales07} used the MS-I to show that the galaxy luminosity function
in fossil groups is consistent with the predictions of a lambda cold dark matter 
universe. Analysing the faint-end slope of the galaxy luminosity function in
observational fossil groups, \cite{proctor11} have observed that the faint 
luminosity tail is well represented by an almost flat slope, suggesting that 
the faint galaxy population is not affected by living in fossil systems.
Nevertheless, most of these works agree that the faint galaxy population
is represented by galaxies mainly down to $-17$ absolute magnitudes, hence, 
a wide range of faint galaxies are out of their analysis.
More recently, an observational work of \cite{lieder13} has attempted to 
gather information about a fainter population of galaxies. These
authors analysed the faint galaxy population of a fossil system down to an absolute
magnitude of $-10.5$ in the R band. They observed that the photometric properties 
of faint galaxies are consistent with those of normal groups or clusters, including 
a normal abundance of faint satellites. 
However, more substantial evidence is needed to confirm these observational
findings. 

\cite{goz14} explored the influence of the faint galaxy population in the formation history of
fossil systems and used the MS-I to study the evolution 
of the luminosity function parameters in fossil and non-fossil systems. 
They confirmed that roughly 80\% of the fossil systems identified 
at early epochs ($z\sim 1$) 
have lost their magnitude gaps before reaching the present time. Analysing the 
faint-end slope of the luminosity function, they observed that there is almost no
evolution of the faint population in fossils, while there is a considerable 
increment of this population in non-fossil systems. 
However, as a result of the nature of the simulation, they considered as faint 
galaxies only those down to $\sim-16$ in the r-band. 

Therefore, to obtain a complete understanding of the evolution of 
the faint galaxy population in fossil groups, a more suitable set of synthetic galaxies is 
needed. Such galaxies can be extracted from the high-resolution N-body 
numerical simulation, the Millennium run simulation II \citep{mII}, 
which is perfect for resolving 
dwarf galaxies using semi-analytic recipes. A particular set of recipes was 
applied to this simulation by \cite{guo11}, producing a highly suitable 
sample of mock galaxies. 
The semi-analytic model has been tuned to reproduce the $z=0$ stellar mass 
function and luminosity function, making it a suitable tool to understand the evolution 
of faint galaxies. 
Therefore, we here use this  publicly 
available tool to study
the evolution of the brightest galaxies in fossil groups from
a semi-analytical point of view and determine  
whether the population of faint galaxies in 
fossil is affected by the formation history of these systems 
compared to the same population in groups considered non-fossils.

The layout of this paper is as follows: 
in Sect.~\ref{glxs} we briefly described the set of semi-analytic galaxies 
used in this work. We identify groups and classify them into 
fossil and non-fossil groups in Sect.~\ref{grps}. 
In Sect.~\ref{brights} we analyse the evolution of the brightest members of 
fossil and non-fossil groups, while 
the selection of the faint population and the analysis of its 
distribution are included in Sect.~\ref{faints}. 
Finally, we summarise our work and discuss the results in Sect.~\ref{theend}.

\section{Mock galaxies}
\label{glxs}
We used a simulated set of galaxies extracted from the  semi-analytic 
model of galaxy formation developed by \cite{guo11}, which has been applied based on the 
Millennium run simulation II \citep{mII}. 

\subsection{N-body simulation}
The Millennium run simulation II is a cosmological tree-particle-mesh 
\citep{xu95} N-body simulation that evolves 
10 billion ($2160^3$) dark matter particles in a 100 $\rm h^{-1} \, Mpc$ periodic 
box, using a comoving softening length of 1 $\rm h^{-1} \, kpc$ \citep{mII}. 
The cosmological 
parameters of this simulation are consistent with WMAP1 data \citep{spergel03}, 
that is, a flat cosmological model with a 
non-vanishing cosmological constant ($\Lambda$CDM): $\Omega_{\rm m}$ =0.25, 
$\Omega_{\rm b}$=0.045, $\Omega_{\Lambda}$=0.75, $\sigma_8$=0.9, $n$=1 and $\rm h$=0.73. 
The simulation was started at $z=127$, with the particles initially positioned 
in a glass-like distribution according to the $\Lambda$CDM primordial 
density fluctuation power spectrum. The mass resolution is 125 times better
than obtained in the Millennium run simulation I \citep{springel05}, which means that 
the mass of each particle is $6.9\times10^6 \, \rm h^{-1} \, \cal{M}_{\odot}$. 
With this resolution, halos of typical dwarf
spheroids are resolved, and halos similar to the mass of our Milky Way have
hundreds of thousands of particles \citep{mII}.

\subsection{Semi-analytic model}
We adopted the simulated set of galaxies built by \cite{guo11}.
This particular semi-analytic model fixed several open questions present in some 
of its predecessors, such as the efficiency of supernova feedback and 
the fit of the stellar mass function of galaxies at low redshifts.
\cite{guo11} also introduced a more realistic treatment 
of satellite galaxy evolution and of mergers, allowing satellites to continue 
forming stars for a longer period of time and reducing the
excessively rapid reddening of the satellite. The model also treats the 
tidal disruption of satellite galaxies. 
Compared to previous versions
of the semi-analytical models, the model of \citeauthor{guo11} has fewer galaxies 
than its predecessors, at any redshifts and in any environment. This is the result of
a stronger stellar feedback that reduces the number of low-mass galaxies, and a model
of stellar stripping, which contributes to reduce the number of intermediate- 
to low-mass galaxies \citep{vulcani14}. 
%%%%%%%%%%%%%%%%%%%%%%%%%%%%%%%%%%%%%%%%%%%%%%%%%%%%%%%%%%%%%%%%%%%%%%%%%%%%%%%
\begin{figure}  
\begin{center}
\includegraphics[width=\hsize]{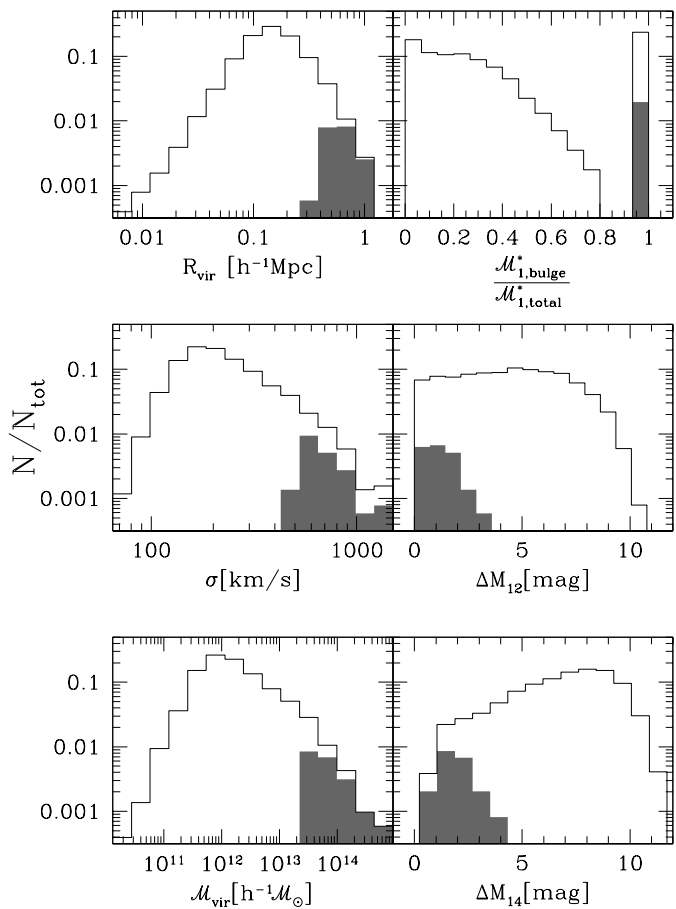}
\caption{Distribution of properties of the semi-analytic galaxy groups: 
3D virial radius 
(top left panel), 3D velocity dispersion (middle left panel), 
virial mass (bottom left panel), 
ratio between stellar mass in bulge and total stellar mass of the 
first-ranked galaxy (top right panel), 
difference between the r-band absolute magnitude of the first- and second-ranked
galaxies within half a virial radius (middle right panel), 
and difference between the r-band absolute magnitude of the first- and 
fourth-ranked
galaxies within half a virial radius (middle right panel).
Empty histograms correspond to FoF groups with 10 or more galaxy members, 
while grey histograms correspond to the 102 FoF groups that have 
virial masses higher than $10^{13.5} \, \rm h^{-1} {\cal M}_\odot$ 
and whose brightest galaxy is an elliptical galaxy 
(${\cal M^{\ast}}_{\rm 1,bulge}/{\cal M^{\ast}}_{\rm 1,total} > 0.7$).
}
\label{f1}
\end{center}
\end{figure}
%%%%%%%%%%%%%%%%%%%%%%%%%%%%%%%%%%%%%%%%%%%%%%%%%%%%%%%%%%%%%%%%%%%%%%%%%%%%%%%
\begin{figure}  
\begin{center}
\includegraphics[width=\hsize]{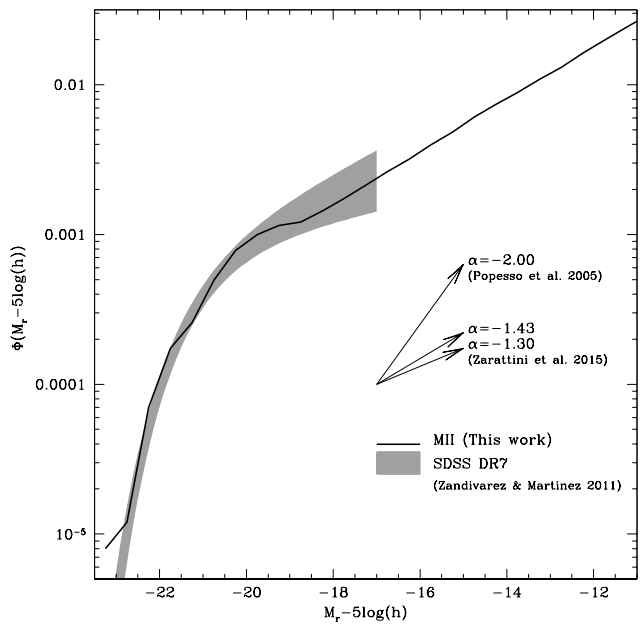}
\caption{Luminosity function of galaxies in groups. The solid line
is the luminosity function for semi-analytical galaxies in groups with 
virial masses higher than $10^{13.5} \, \rm h^{-1} {\cal M}_\odot$ 
and whose brightest galaxy is an elliptical galaxy.
Grey region show the results for galaxies in groups in the SDSS DR7
obtained by \cite{zandivarez11}. Upper and lower arrows represent the 
faint end slope of the luminosity function obtained by \cite{popesso05} and 
\cite{fogo5}, while middle arrow correspond to the value obtained in this work.
}
\label{f1_bis}
\end{center}
\end{figure}
%%%%%%%%%%%%%%%%%%%%%%%%%%%%%%%%%%%%%%%%%%%%%%%%%%%%%%%%%%%%%%%%%%%%%%%%%%%%%%%

This model produces a complete sample when considering galaxies with 
stellar masses higher than $\sim 10^{6.4} \, \rm h^{-1} \, \cal{M}_{\odot}$. This implies
that the galaxy sample is almost complete down to an absolute magnitude in
the $r_{SDSS}$-band of -11. 

Since different cosmological parameters have been found from WMAP7 \citep{komatsu11}, 
it might be argued that the studies carried out in the present simulation 
 produce results that do not agree with the current cosmological model. 
However, \cite{guo13} have 
demonstrated that the abundance and clustering of dark halos and galaxy properties, 
including clustering, in WMAP7 are very similar to those found in WMAP1 for $z\leq3$, 
which is far inside the redshift range of interest in this work.

\section{Group samples}
\label{grps}
\subsection{Identification of friends-of-friends groups}
\label{FoF}
Groups of galaxies were identified by using a friends-of-friends (FoF) algorithm in real
space \citep{davis85} applied to the mock galaxies in the simulation box.

To study different evolutionary stages of the simulated groups, we performed nine
identifications in different outputs, from redshift $z=0$ to $z=1.08$ 
($\sim 8 \, \rm Gyr$), each output spaced by $\sim 0.1$. 
Following \cite{fof14}, we considered
that the linking length \textit{b} used by the algorithm
 to cluster galaxies depends on the redshift 
as follows:
%%%%%%%%%%%%%%%%%%%%%%%%%%%%%%%%%%%%%%%%%%%%%%%%%%%%%%%%%%%%%%%%%%%%%%%%%%%%%%%%%%%%%%%%%
\begin{equation*}\label{eq:b} 
b(z)=b_{0}\left(0.24\frac{\Delta_{\rm vir}(z)}{178}+0.68\right)^{-1/3}
\end{equation*}
%%%%%%%%%%%%%%%%%%%%%%%%%%%%%%%%%%%%%%%%%%%%%%%%%%%%%%%%%%%%%%%%%%%%%%%%%%%%%%%%%%%%%%%%%
where the enclosed overdensity of haloes, $\Delta_{\rm vir}$, depends on the cosmology and 
the value of redshift according to
%%%%%%%%%%%%%%%%%%%%%%%%%%%%%%%%%%%%%%%%%%%%%%%%%%%%%%%%%%%%%%%%%%%%%%%%%%%%%%%%%%%%%%%%%
\begin{equation*}\label{eq:deltavir}
\Delta_{\rm vir}(z)=18\pi^{2}\left[1+0.399\left(\frac{1}{\Omega_{\rm m}(z)}-1\right)^{0.941}\right]
\end{equation*} 
%%%%%%%%%%%%%%%%%%%%%%%%%%%%%%%%%%%%%%%%%%%%%%%%%%%%%%%%%%%%%%%%%%%%%%%%%%%%%%%%%%%%%%%%%
where
$\left(\frac{1}{\Omega_{\rm m}(z)}-1\right)=\left(\frac{1}{\Omega_{0}}-1\right)(1+z)^{-3}$
and $\Omega_0$ is the dimensionless matter density parameter at the present.
We identified groups in a galaxy catalogue instead of in one
of dark matter particles. Therefore, and following previous studies 
\citep{eke04,berlind06,fof14}, 
we used a fiducial linking length of $b_0=0.14$ 
(instead of the conventional $b_0=0.2$ for DM halos), which corresponds to a contour
overdensity contrast of $\sim 433$. 
Using this prescription, we obtained a galaxy group catalogue at $z=0$ 
comprising 5116 systems with ten or more galaxy members. 

In Fig.~\ref{f1} we show the distributions
of the physical properties of the simulated galaxy groups identified at $z=0$ 
(empty histograms). 
In the left column, from top to bottom, 
we show the 3D virial radius, the 3D velocity dispersion,
and the group virial mass.
The 3D virial radius was computed according to
%%%%%%%%%%%%%%%%%%%%%%%%%%%%%%%%%%%%%%%%%%%%%%%%%%%%%%%%%%%%%%%%%%%%%%%%%%%%%%%%%%%%%%%%%
\begin{equation*}\label{eq:rvir} 
R_{\rm vir}=\frac{N_{\rm g}(N_{\rm g}-1)}{2\sum_{i}\sum_{j<i}{(r_{ij})^{-1}}}
\end{equation*}
%%%%%%%%%%%%%%%%%%%%%%%%%%%%%%%%%%%%%%%%%%%%%%%%%%%%%%%%%%%%%%%%%%%%%%%%%%%%%%%%%%%%%%%%%
where $N_{\rm g}$ is the total number of galaxy members and $r_{ij}$ are the
3D intergalaxy separations.
The 3D group velocity dispersion, $\sigma$, was estimated as the root 
mean square velocity dispersion from the group galaxy members. And finally, 
the virial mass was computed as follows:
%%%%%%%%%%%%%%%%%%%%%%%%%%%%%%%%%%%%%%%%%%%%%%%%%%%%%%%%%%%%%%%%%%%%%%%%%%%%%%%%%%%%%%%%%
\begin{equation*}\label{eq:mvir} 
{\cal M}_{\rm vir}=\frac{\sigma^{2}\, R_{\rm vir}}{\rm G}
\end{equation*}
%%%%%%%%%%%%%%%%%%%%%%%%%%%%%%%%%%%%%%%%%%%%%%%%%%%%%%%%%%%%%%%%%%%%%%%%%%%%%%%%%%%%%%%%%
where $G$ is the gravitational constant.
The sample of groups has median virial radius of $ 0.14 \, \rm h^{-1} \, Mpc$, median velocity dispersion of $ 200 \, \rm km/s$, and median virial mass of $ 1.3\times10^{12} \, \rm h^{-1} \, {\cal M}_\odot$. 

In the right column of Fig.\ref{f1} we show the distributions of properties that are
used in the following sections to select fossil and non-fossil systems (empty histograms).
%\textcolor[rgb]{1,0.501961,0}{ 
%From top to bottom:
%ratio of the mass of stars in bulge and the total stellar mass of the first ranked galaxy 
%of the groups, difference between the r-band absolute magnitude of the first and second 
%ranked galaxies within half the virial radius of the groups, 
%and difference between the r-band absolute magnitude of the first and fourth 
%ranked galaxies within half the virial radius of the groups.\LEt{this
%simply repeats the information of the figure caption, please
%remove from the main text}} 

%%%%%%%%%%%%%%%%%%%%%%%%%%%%%%%%%%%%%%%%%%%%%%%%%%%%%%%%%%%%%%%%%%%%%%%%%%%%%%%
\begin{figure*} 
\begin{center}
\includegraphics[width=9cm]{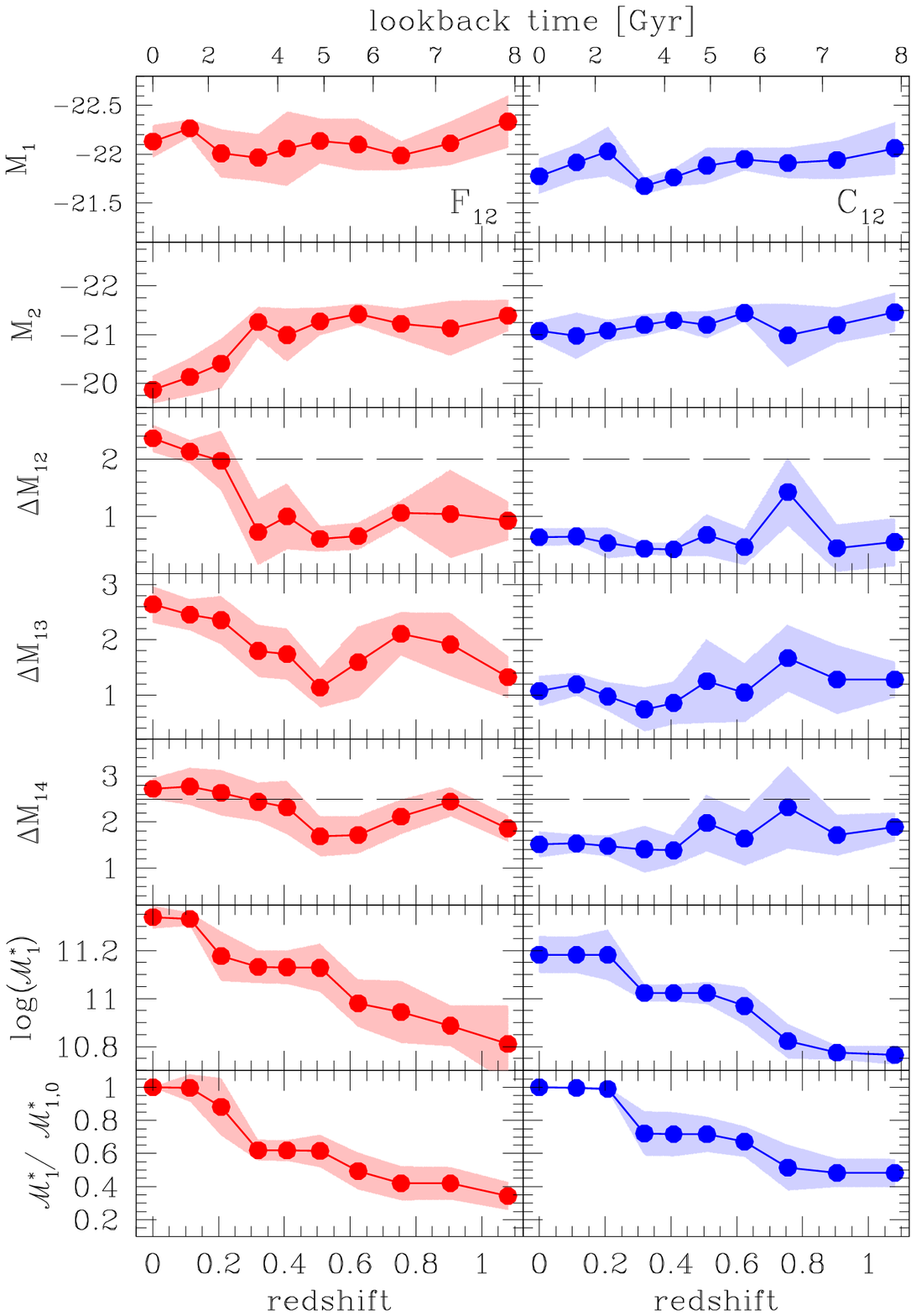}
\includegraphics[width=9cm]{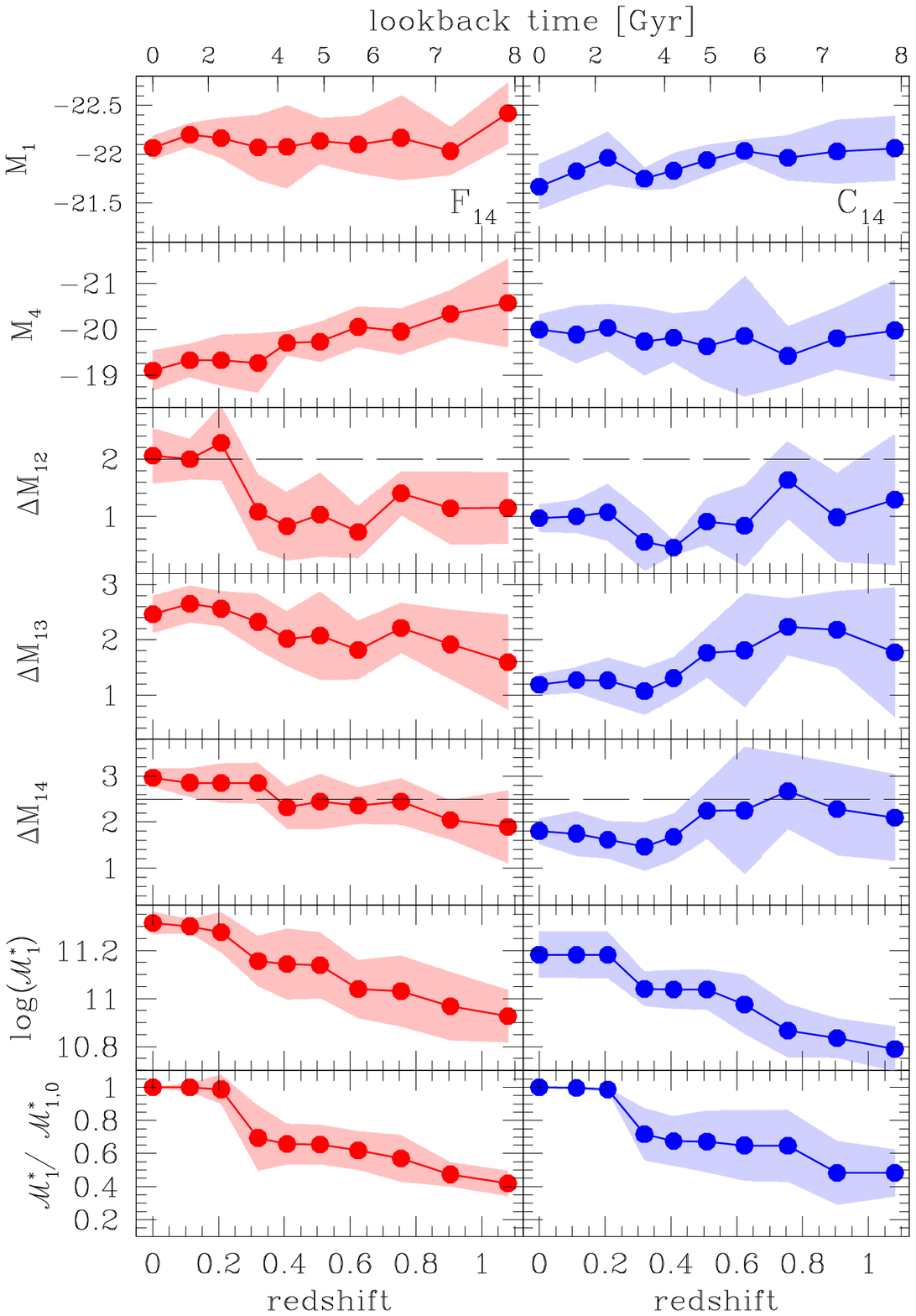}
\caption{Evolution of the median properties of fossil and control groups.
The left plots correspond to the sample of fossils and controls defined using the $\Delta M_{12}$ magnitude gap, while the right plots correspond 
to the definition based on the $\Delta M_{14}$ magnitude gap. 
Panels in the first row show the evolution of the r-band absolute magnitude 
of the brightest galaxy of the groups; 
the second row shows the absolute magnitude of the 
second and fourth brightest galaxies;
the third, fourth, and fifth rows show 
the evolution of the r-band absolute magnitude gap 
between the brightest and the 
second, third, and fourth brightest galaxies, respectively, 
the dashed horizontal lines correspond to the thresholds used 
to define fossil groups;
the sixth row shows the stellar mass of the brightest galaxy, 
and the last row shows the evolution of the stellar mass of 
the brightest galaxy normalised to its final stellar mass.
Error bands are the semi-interquartile ranges.
}
\label{f2}
\end{center}
\end{figure*}
%%%%%%%%%%%%%%%%%%%%%%%%%%%%%%%%%%%%%%%%%%%%%%%%%%%%%%%%%%%%%%%%%%%%%%%%%%%%%%%

\subsection{Fossil groups}
\label{fossilsec}
\cite{jones03} identified fossil groups as spatially extended X-ray sources with an X-ray
luminosity $L_{x} \geq 10^{42} \rm \, h_{50}^{-2} \, erg/s$, whose optical counterpart
is a bound system of galaxies with $\Delta M_{12}\geq 2$, where $\Delta M_{12}$
is the difference in absolute magnitude in the R band between the brightest and the second
brightest galaxies located within half the project virial radius of the systems.
Using this definition, it is assumed that galaxies within 
half the virial radius have had time to merge within a Hubble time, and also that normal 
elliptical galaxies that are not located at the centre of the groups will not be 
chosen as potential fossil groups \citep{lieder13}.
In addition to the conventional criteria, there exists
an alternative criterion developed by \cite{dariush10}. These authors
found that imposing the magnitude gap in the R band between the brightest and the
fourth brightest galaxies within half the projected virial radius to be larger than 
$2.5$ magnitudes, $\Delta M_{14}\geq 2.5$, identifies $50\%$ 
more early-formed systems, and such systems, on average, retain their fossil phase longer. 
However, the conventional criteria perform marginally better at finding early-formed 
groups at the high-mass end of the virial mass distribution of groups.
In this work, we used the two criteria defined above to identify fossil groups with 
the purpose of performing comparative studies. 
 
Given that we do not have X-ray luminosity in the simulation boxes, 
we adopted a lower cut-off in group virial masses, 
${\cal M}_{\rm vir} \geq 10^{13.5}\, \rm h^{-1} \, {\cal M}_{\odot}$ 
 to maximise the probability that the selected systems 
are strong X-ray emitters \citep{dariush07}. 
Moreover, we included a criterion to ensure that the brightest galaxy of the 
selected groups is elliptical, 
as is found in all the observational fossil groups known to date. 
Following \cite{bertone07}, we classified as ellipticals those galaxies whose
ratio between the stellar mass of the bulge and the total stellar mass is 
higher than a given threshold: 
${\cal M^{\ast}}_{\rm bulge}/{\cal M^{\ast}}_{\rm tot} > 0.7$. The distributions of
properties of the $102$ FoF groups that satisfy these two criteria are shown
as grey histograms in Fig.~\ref{f1}.
We also compare the distribution of r-band absolute magnitudes of galaxies in 
these simulated groups with results from observations in Fig.~\ref{f1_bis}. The grey region was built from the 
best-fitting Schechter parameters of the luminosity function of galaxies in groups identified in 
the SDSS DR7 by \cite{zandivarez11}\footnote{We introduced a shift in the $M^*$ parameter to 
account for the shift to z=0.1 that these authors used. According to \cite{blanton05}, 
$r=r^{0.1}-0.22$}. These fits were obtained only for galaxies brighter than 
$M_r-5\log(\rm h)=-17$. 
The lower envelope corresponds to groups with masses $\sim 10^{13.5}$, while 
the upper envelope corresponds to groups with masses higher than $10^{14.1}$. The bright end of the luminosity function of the semi-analytical galaxies in groups agrees with the observations. 
The behaviour of the faint-end slope of the luminosity function was compared with the observational 
results obtained by \cite{popesso05} and \cite{fogo5} (arrows in Fig.~\ref{f1_bis}).
The slopes obtained by these authors encompass
the value obtained in this work of $\alpha \simeq -1.43$.

Summarising, the criteria applied to the simulated groups to select fossil 
groups according to the two definitions are 
\begin{itemize}
\item fossil $F_{12}$: ${\cal M}_{\rm vir} \geq 10^{13.5}\, \rm h^{-1}\, {\cal M}_{\odot}$; 
${\cal M^{\ast}}_{\rm 1,bulge}/{\cal M^{\ast}}_{\rm 1,tot} > 0.7$ ; 
and $\Delta M_{12}\geq 2$ \citep{jones03}, and 
\item fossil $F_{14}$: ${\cal M}_{\rm vir} \geq 10^{13.5}\, \rm h^{-1}\, {\cal M}_{\odot}$; 
${\cal M^{\ast}}_{\rm 1,bulge}/{\cal M^{\ast}}_{\rm 1,tot} > 0.7$ ; 
and $\Delta M_{14}\geq 2.5$  \citep{dariush10}.
\end{itemize}

We selected 14 groups as $F_{12}$ and 22 groups as $F_{14}$.  
Finally, we examined the evolution of these fossil groups at different times. 
We identified normal groups in each previous snapshot (see Sect.~\ref{FoF}). 
Then, we selected those groups that hosted the progenitor of the brightest galaxy 
of the fossil group at z=0.  
Since fossil groups are considered undisturbed and old systems, 
we only considered those systems that have already assembled more 
than the 50 per cent of their final mass at $z\gtrsim 0.8$. 
The samples comprise 9 fossil groups in the $F_{12}$ category and 15 
in the $F_{14}$ category. 
We note that $78\%$ of the $F_{12}$ are also included in the $F_{14}$ sample, in agreement with the results of \cite{dariush10} (75\%). 
Conversely, $47\%$ of the $F_{14}$ satisfy the $\Delta M_{12}$ criterion; 
this percentage is higher than found by \citeauthor{dariush10} (35\%). 
This difference may be a consequence of the several restrictions 
imposed on our sample selection that have not been considered by these authors.
Finally, we discarded 3  $F_{14}$ groups to build a control sample that matches
 the  distribution of their virial masses (see Sect.~\ref{control} for details). 
The final $F_{14}$ sample comprises 12 groups. 

\subsection{Control groups}
\label{control}
To perform a comparative study, we selected samples of non-fossil groups. 
These samples satisfy 
 
\begin{itemize}
\item non-fossil $nF_{12}$: ${\cal M}_{\rm vir} \geq 10^{13.5}\, \rm h^{-1}\, {\cal M}_{\odot}$; ${\cal M^{\ast}}_{\rm 1,bulge}/{\cal M^{\ast}}_{\rm 1,tot} > 0.7$ ; and $\Delta M_{12} < 2$, and
\item non-fossil $nF_{14}$: ${\cal M}_{\rm vir} \geq 10^{13.5}\, \rm h^{-1}\, {\cal M}_{\odot}$; ${\cal M^{\ast}}_{\rm 1,bulge}/{\cal M^{\ast}}_{\rm 1,tot} > 0.7$ ; and $\Delta M_{14} < 2.5$.
\end{itemize}

We selected 88 groups as $nF_{12}$ and 80 groups as $nF_{14}$. 
We also examined the evolution of
the mass assembly of these groups and saved only those groups that assembled half of their 
final virial mass at redshifts higher than $z\simeq 0.8$. 
This criterion restricted the samples to 38 $nF_{12}$ and 34 $nF_{14}$. 

Since cluster formation histories depend strongly on the mass of the systems,
differences in the virial mass distributions of the fossil and non-fossil samples could
introduce biases into the results. Therefore, we need subsamples of non-fossil
groups that reproduce the virial mass distribution of their corresponding fossil 
groups for each category (``$12$'' or ``$14$'').  
After performing a two-sample comparison between the distributions of virial masses of 
fossil samples and $50$ random selections of non-fossil samples, 
we selected the non-fossil group samples 
with the highest probability values from the two-sample Kolmogorov-Smirnov (KS) test.
Moreover, given that the definition of fossil groups can be thought of as a tail of the 
distribution of differences in magnitudes between the brightest galaxies, 
we also included a similar restriction in the magnitude gap of non-fossil groups 
to select the opposite tail of the distribution of gaps. 
The upper threshold imposed to limit the magnitude gaps of control groups was chosen to obtain samples with a similar number of groups as their counterpart fossil samples. 
Then, the final control group samples have to fulfil the following
criteria: 
\begin{itemize}
\item Control $C_{12}$: ${\cal M}_{\rm vir} \geq 10^{13.5}\, \rm h^{-1}\, {\cal M}_{\odot}$; ${\cal M^{\ast}}_{\rm 1,bulge}/{\cal M^{\ast}}_{\rm 1,tot} > 0.7$ ; KS probability $> 0.95$ and $\Delta M_{12} < 0.9$.
\item Control $C_{14}$: ${\cal M}_{\rm vir} \geq 10^{13.5}\, \rm h^{-1}\, {\cal M}_{\odot}$; ${\cal M^{\ast}}_{\rm 1,bulge}/{\cal M^{\ast}}_{\rm 1,tot} > 0.7$ ; KS probability $> 0.95$ and $\Delta M_{14} < 2 $.
\end{itemize}

We obtained the final samples of control groups that comprise 9 $C_{12}$ and 
12 $C_{14}$. 
In this way, we have constructed samples of similar numbers of fossil and control groups 
that have the same mass distribution, similar assembly times (old systems), 
but some (fossils) have been able to develop a large magnitude gap between 
their brightest galaxies, while the others (controls) do not. 

\section{Brightest galaxies}
\label{brights}

Figure \ref{f2} shows the evolution of the median of different properties of the final samples of fossil (left columns)
and control (right columns) groups as a function of time (redshift) 
for samples defined using the
$\Delta M_{12}$ criterion (left plots) and the $\Delta M_{14}$ criterion
(right plots).
%\LEt{this is again repetition of the figure caption,
%please remove} \textcolor[rgb]{1,0.501961,0}{Error bars are the semi-interquartile ranges for each snapshot.
%From top to bottom, we show the trends for 
%the r-band absolute magnitude of the brightest galaxy of the groups ($M_1$),
%the r-band absolute magnitude of the second and fourth brightest galaxy within half 
%the virial radius of the groups ($M_2$ or $M_4$),
%the absolute magnitude gap within half the virial radius between
%the brightest and the second ($\Delta M_{12}$), 
%the third ($\Delta M_{13}$), 
%and the fourth ($\Delta M_{14}$) brightest galaxies,
%the stellar mass of the brightest galaxy, 
%and the stellar mass of the brightest galaxy normalised 
%to its final stellar mass (${\cal M}_1^{\ast}/{\cal M}_{1,0}^{\ast}$).} 

The first-ranked galaxies in fossil systems are typically
brighter than their counterparts in control groups of the same virial mass and shows little or no evolution with time. 
On the other hand, the magnitude
of the second- or fourth-ranked galaxies of fossil groups shows evolution with 
time (probably being replaced) and produces a change in the magnitude gap that is used to 
split between fossils and non-fossils (this is better seen in the $F_{12}$ sample).
The magnitude gap used to define a group as fossil is above the fixed threshold  
of around $z\simeq 0.2$ for $F_{12}$ (left plot, third row), 
and around $z\simeq 0.3$ for $F_{14}$ (right plot, fifth row), 
although many $F_{14}$  have fulfilled the fossil 
criterion for a longer period of time (since $z\sim 0.8$). 
This means that on average, fossil systems have reached their status during 
the past $3.5$ Gyr.
Moreover, we found that the control groups $C_{12}$ have not experienced 
a fossil phase during their whole history, while some of the $C_{14}$ have had a 
fossil phase in the earlier stages of group formation.

We also analysed the evolution of the magnitude gap between the
brightest and the second, third, and fourth brightest galaxies in each group sample. 
In the $F_{12}$ sample,
for $z \gtrsim 0.3$ the magnitude gap between the brightest and the 
second brightest galaxies is
$\Delta M_{12} \lesssim 1$, while $\Delta M_{13}$ and $\Delta M_{14}$
are $\sim 2$. Hence, fossil groups at earlier times had two dominant galaxies 
with similar absolute magnitudes surrounded by much fainter neighbours. 
For $z \lesssim 0.2$, all the magnitude gaps shown in this 
figure increase, and the change in $\Delta M_{12}$ is more noticeable. 
We infer that there has been a mayor event between the first- and second-ranked galaxies, and as a consequence, 
there has been a rearrangement in the luminosity ranking 
leading to the observed changes in the magnitude gaps. 
In the $C_{12}$ groups there were no two dominant galaxies during the past 8 Gyr.  
The difference in absolute magnitude between one galaxy and the next in the luminosity
ranking is always $\sim 0.5$ magnitudes. 
A similar analysis can be performed in the $F_{14}$ and $C_{14}$ samples.

Analysing the stellar mass of the first ranked galaxies, it can be seen that  
it is higher in fossil groups than in control groups in the whole redshift range. 
At redshift zero, the brightest galaxy in fossils is $\sim 50\%$ more massive 
than in control groups. 
Moreover, the brightest galaxies in the $F_{12}$ have reached $\sim 50\%$ 
of their final mass at $z\simeq 0.6$, 
while the brightest galaxies of control $C_{12}$ have assembled $\sim 50\%$ 
of their final mass before $z\simeq 1$. 
This difference does not exist when analysing the $F_{14}$ sample, since the brightest galaxies
in these systems have assembled their mass on average at redshifts as high as 
in the control samples.  
From these panels, we observe that the stellar mass of 
the brightest galaxy of fossil and control groups shows a
noticeable increment ($\gtrsim 20\%$) around $z\simeq 0.3$, 
regardless of how the samples are defined. 
This result indicates a major event in all the group samples.  
As a result of the different galaxy luminosity sampling of the groups, 
this event was also inferred when analysing the magnitude gaps for fossil groups, but it
was not observable from the magnitude gaps for control groups. 

Fossil groups - regardless of the criterion adopted to define them - are therefore characterised not only by their early formation times, which can
also be achieved by control groups, but it is also necessary that at high redshifts two similar massive galaxies exist at 
their cores,  while the next-ranked galaxies are much fainter. 
Whether a group becomes fossil is therefore defined at the very 
beginning of the group history  by its luminosity content.
 
%%%%%%%%%%%%%%%%%%%%%%%%%%%%%%%%%%%%%%%%%%%%%%%%%%%%%%%%%%%%%%%%%%%%%%%%%%%%%%%
\begin{figure} 
\begin{center}
\includegraphics[width=9cm]{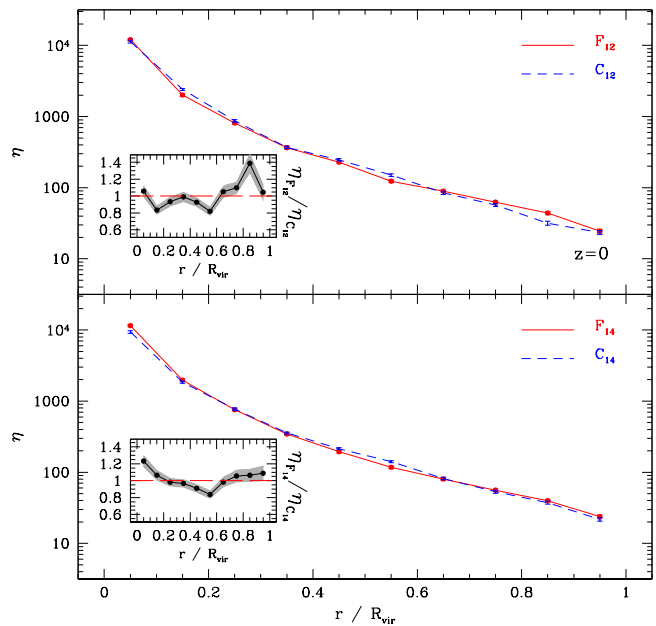}
\caption{Number density profiles of faint galaxies in fossil and 
control groups. 
The upper panel shows the number density of faint 
galaxies in fossils $F_{12}$ and control $C_{12}$ groups
as a function of the 3D normalised distance to the centre of the groups. 
The lower panel shows the same situation as in the top panel for fossils $F_{14}$
and control $C_{14}$ groups. Error bars are computed using the bootstrap resampling technique. 
Inset panels show the ratio between the number density profiles in fossils and controls. 
Error bars are computed via error propagation.}
\label{f5}
\end{center}
\end{figure}
%%%%%%%%%%%%%%%%%%%%%%%%%%%%%%%%%%%%%%%%%%%%%%%%%%%%%%%%%%%%%%%%%%%%%%%%%%%%%%%

\label{evol} 
\begin{table}
\centering
\caption{Number of faint galaxies in the composite clusters at 
each different evolutionary stage \label{n_faints}}
\begin{tabular}{|c|cc|cc|}
redshift & $F_{12}$ & $C_{12}$ & $F_{14}$ & $C_{14}$ \\
\hline
0.000 & 4461 & 4552 & 5222 & 5503 \\
0.116 & 4143 & 4248 & 5095 & 5158 \\
0.208 & 3823 & 4129 & 4693 & 5070 \\
0.320 & 3452 & 3872 & 4279 & 4786 \\
0.408 & 3274 & 3855 & 4138 & 4649 \\
0.509 & 3456 & 3766 & 4187 & 4325 \\
0.624 & 3528 & 3633 & 4194 & 3944 \\
0.755 & 3224 & 3175 & 3819 & 3661 \\
0.905 & 2800 & 2666 & 3303 & 3039 \\
1.080  & 2431 & 2304 & 2778 & 2334 \\ 
\hline
\end{tabular}
\end{table}

%%%%%%%%%%%%%%%%%%%%%%%%%%%%%%%%%%%%%%%%%%%%%%%%%%%%%%%%%%%%%%%%%%%%%%%%%%%%%%%
\begin{figure*} 
\begin{center}
\includegraphics[width=9cm]{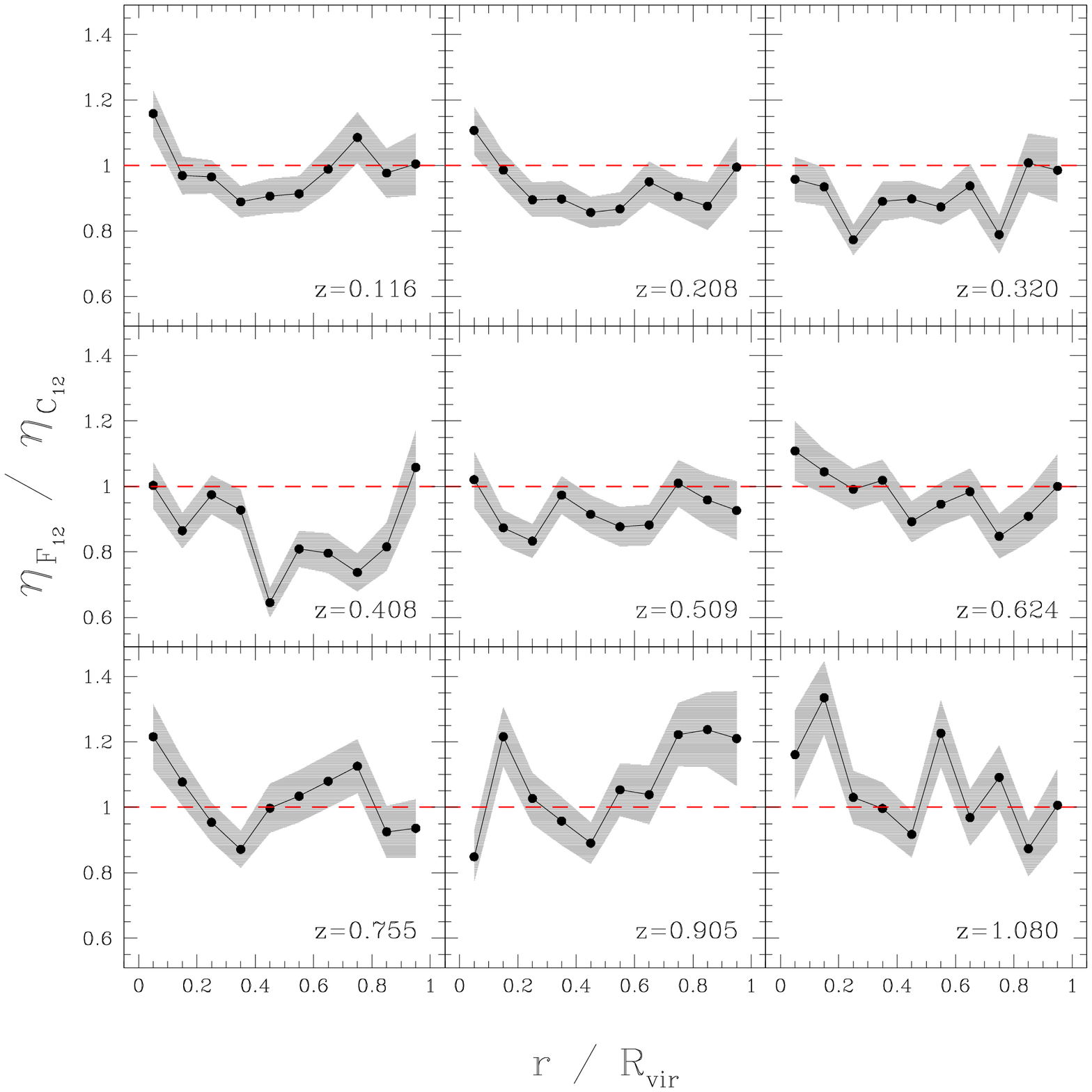}
\includegraphics[width=9cm]{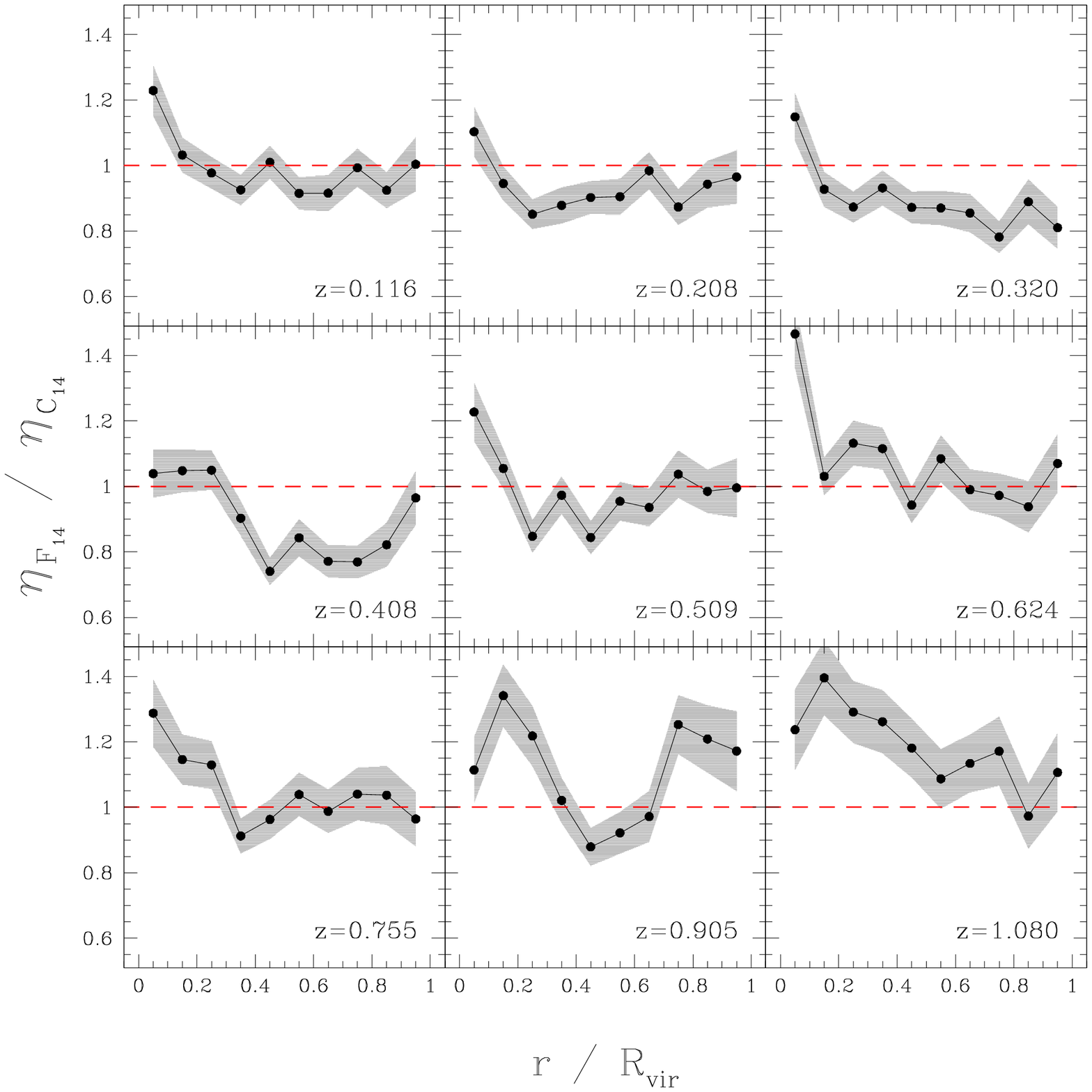}
\caption{Ratios of number density profiles of faint galaxies in fossil and control 
groups at different evolutionary stages. 
The left plot shows the evolution of
the ratio between the number density of faint galaxies in fossil
$F_{12}$ and control $C_{12}$ groups as a function of the 3D normalised
distance to the centre of the group. The right plot shows the same, but
using fossils $F_{14}$ and control $C_{14}$.
}
\label{f6}
\end{center}
\end{figure*}
%%%%%%%%%%%%%%%%%%%%%%%%%%%%%%%%%%%%%%%%%%%%%%%%%%%%%%%%%%%%%%%%%%%%%%%%%%%%%%%

\section{Faintest galaxies}
\label{faints}
To study the faint galaxy population in fossil and 
control groups, we selected from each system those member galaxies that lie within
one virial radius of the centre of the group, whose r-band absolute 
magnitudes are in the range $-16\le M_r-5\log(\rm h)\le-11$. Then we analysed the distribution of these
galaxies around the group centres. This selection was also performed in the different
evolutionary stages analysed in this work (see Sect.~\ref{FoF}).  

To increase the statistical significance of the results, 
we combined in each snapshot all groups of each category ($F_{12}$, $C_{12}$, $F_{14}$, $C_{14}$)
 to produce composite clusters formed with 
faint galaxies, properly scaled to take into account the different sizes of groups 
within each category.
The number of faint galaxies that build the composite clusters in the different 
snapshots are quoted in Table~\ref{n_faints}.

The centre of each individual group was defined as the position of the brightest galaxy
of the group, and the distances of each faint galaxy to the centre were normalised to the 
3D virial radius of its host group. Then, the number density profile of the 
composite cluster was computed as a function of the normalised groupcentric distances 
($\eta(r/R_{\rm vir})$).

In Fig.~\ref{f5} we show the number density profiles of faint galaxies 
in fossil and control groups at $z=0$. 
%}\LEt{this is again a
%repetition, please remove}
%\textcolor[rgb]{1,0.501961,0}{The upper panel corresponds to the profiles 
%when the $\Delta M_{12}$ is used to define fossils or controls, while in the lower 
%panel we shown the same but for the samples defined using the $\Delta M_{14}$ criterion. 
%Error bars for each number density profile were computed using 
%the bootstrap resampling technique. In the inset panels we show the ratio 
%between the profiles of fossils and control groups, 
%the error bars in these panels were computed 
%using the error propagation formula.}
We observe from these panels that on average, the number density 
profiles of faint galaxies around fossils and controls span 
the same range of densities and, apart from local fluctuations, a similar distribution
of faint galaxies is observed, 
regardless of the magnitude criterion adopted to define the samples. 

%%%%%%%%%%%%%%%%%%%%%%%%%%%%%%%%%%%%%%%%%%%%%%%%%%%%%%%%%%%%%%%%%%%%%%%%%%%%%%%
\begin{figure*}  
\begin{center}
\includegraphics[width=9cm]{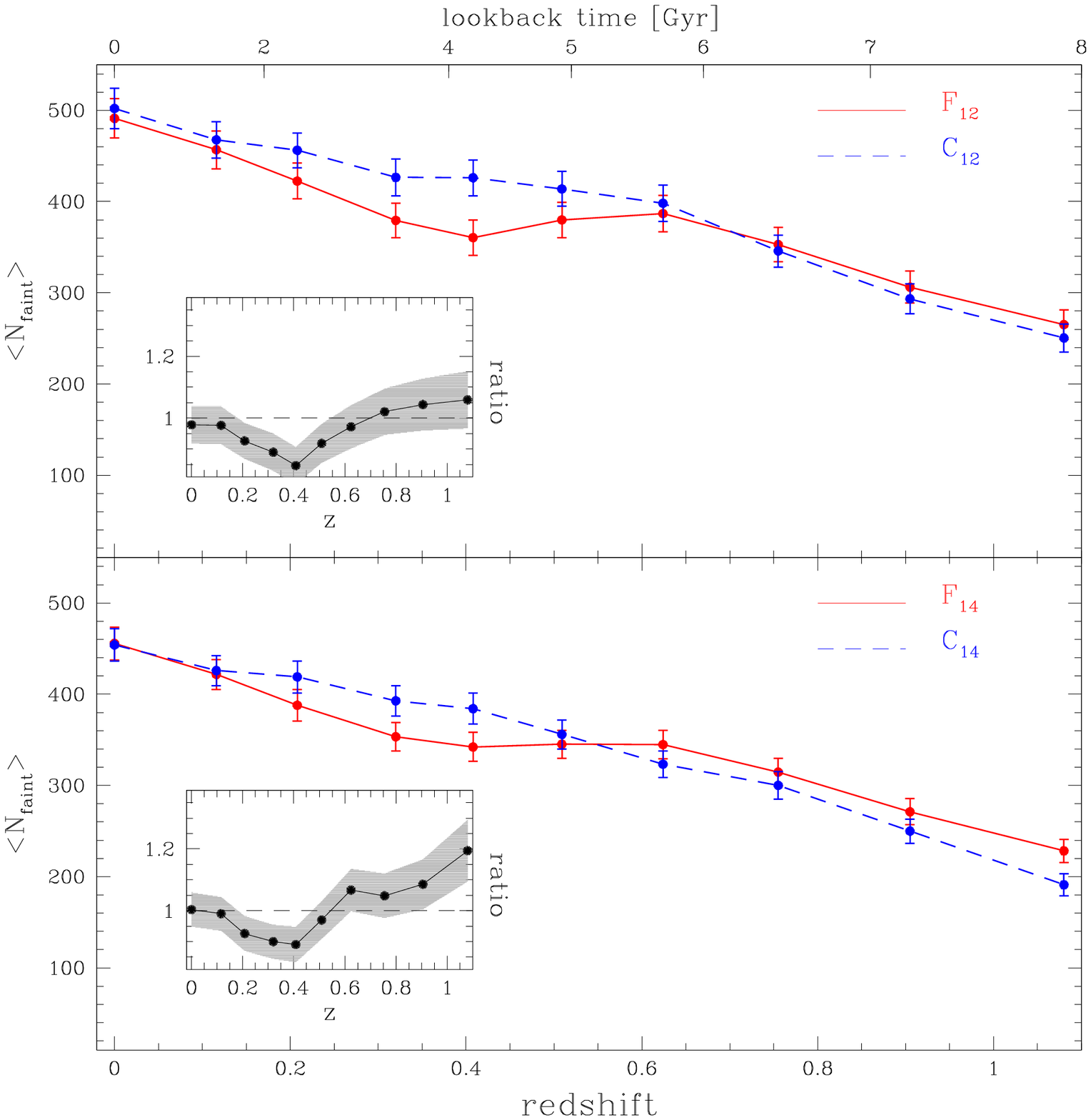}
\includegraphics[width=9cm]{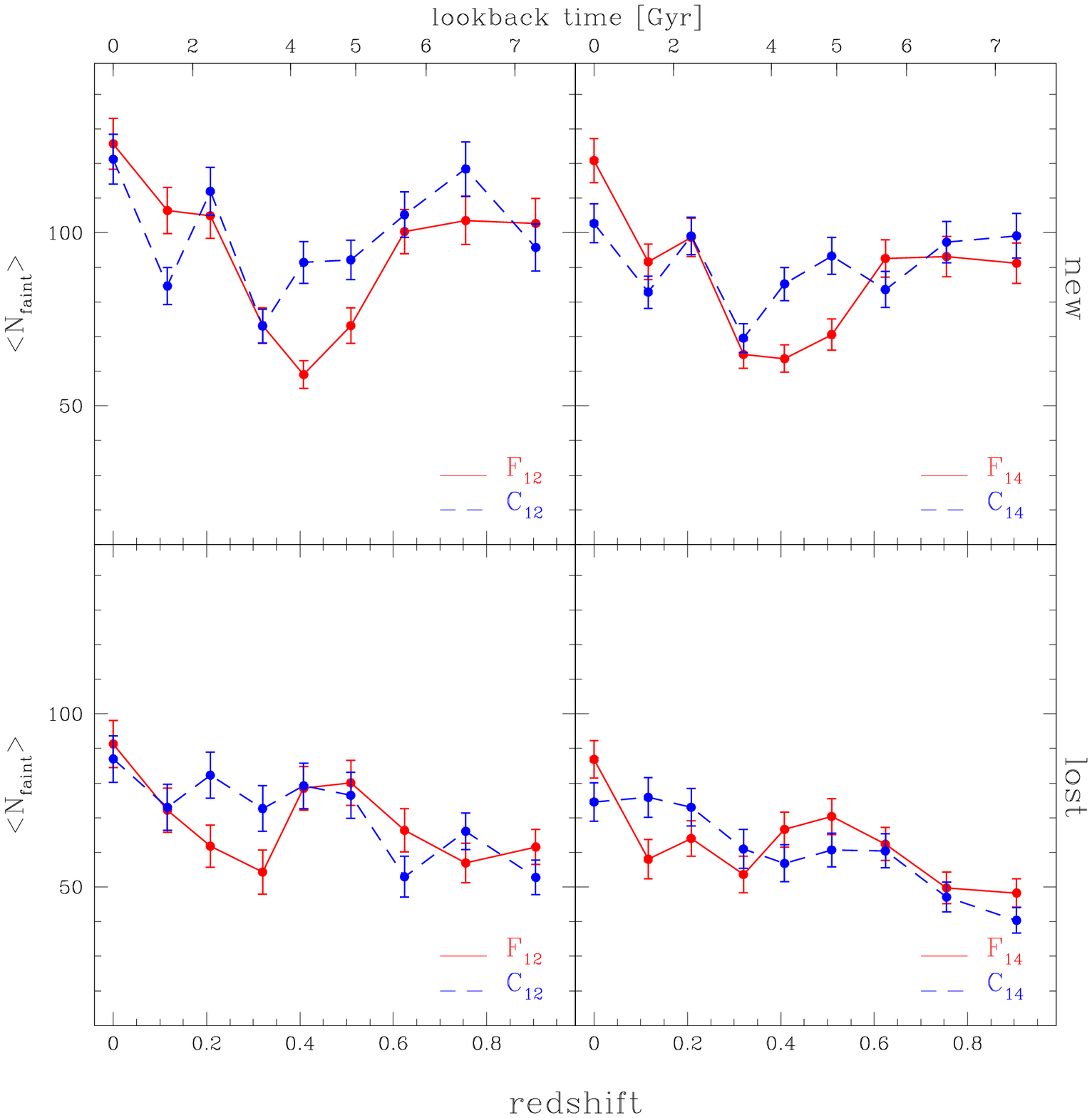}
\caption{Left plot: Mean number of faint galaxies in fossils (solid lines) 
and controls (dashed lines) as a function of redshift.
The upper panel shows the evolution observed in groups defined using 
the $\Delta M_{12}$ criterion, while lower panel corresponds to groups 
identified according to the $\Delta M_{14}$ criterion. 
Error bars are computed using the bootstrap resampling technique. 
The inset panels show the ratio 
between the mean number of faint galaxies in fossil and control
groups.
Right plot: Mean number of faint galaxies classified as new members (upper panels) or lost members (lower panels).
Error bars are computed using the error propagation formula }
\label{f4}
\end{center}
\end{figure*}
%%%%%%%%%%%%%%%%%%%%%%%%%%%%%%%%%%%%%%%%%%%%%%%%%%%%%%%%%%%%%%%%%%%%%%%%%%%%%%%

It is interesting to compare the behaviour of the faint galaxy population in fossils 
and controls throughout the history of the galaxy systems. 
We computed the number density profiles of faint galaxies at different evolutionary stages. 
Figure \ref{f6} shows the evolution of the ratios between the number density profiles 
of faint galaxies in fossil and control groups of each category. 
The population of faint galaxies within
fossil and control groups behave differently during their evolution 
compared with the behaviours observed for $z=0$.
The left plot of Fig.~\ref{f6} (using $\Delta M_{12}$, \citealt{jones03}) 
shows that for $z\ge 0.755$ the number density of faint galaxies in fossil groups 
was higher than observed in control groups in most of the range of normalised 
distances. Later, in the range $ 0.3\lesssim z \lesssim 0.5$, 
the opposite behaviour is observed: 
control groups are denser in faint galaxies than fossil groups in the whole range of distances. 
For the latest times ($z\lesssim 0.2$), the distribution of faint galaxies 
in fossils and non-fossils
tends to behave more similarly.
In the right plot of Fig.~\ref{f6} (using $\Delta M_{14}$, \citealt{dariush10}) 
we observe a very similar general behaviour as the previously described for the 
left panel, but with a tendency to show a higher density in the central regions 
of fossil groups than in their controls.

To better understand the evolution of the faint galaxy population, 
we show in the left plot of Fig.~\ref{f4} the mean number of faint galaxies in fossils and controls as a function of redshifts. 
This number has been computed from the number density profiles as
$\langle N_{faint} \rangle = N_G^{-1} \, \int_0^1 \eta \left(r/R_{\rm vir}\right) dV  $, 
where dV is a volume differential, and $N_G$ is the number of groups in each sample. 
The upper panel corresponds to groups defined with the 
$\Delta M_{12}$ criterion, while groups identified with the $\Delta M_{14}$ criterion are
shown in the lower panel.
Starting at redshift $z\simeq1$, there are slightly more faint galaxies in fossil groups. In the range $0.3 \lesssim z \lesssim 0.6$, 
control groups present more faint galaxies. 
Finally, the number of faint galaxies become similar in 
both group samples towards redshift zero. 
These results are also observed in the inset panels, 
where the ratios between the mean number of faint galaxies in fossil and control groups
for each category are shown. This behaviour is observed for both categories of fossil groups.

This clearly shows that the frequency evolution
of faint galaxies 
for fossil and control groups is quite different. On one hand,  
the number of faint galaxies in control groups smoothly increases 
from $z\simeq 1$ to $z=0$. 
On the other hand, there is an abrupt change in the evolution of the number 
of faint galaxies in fossils in the range $0.3 < z < 0.6$: 
the number of faint galaxies in fossil
groups ceases to increase, showing an almost constant slope. 
For $z\lesssim 0.3$ the number of faint galaxies in fossils rapidly grows until 
it finally matches the number of faint galaxies in control groups at present. 

This different behaviour between fossils and controls 
could be caused by different accretion rates of faint galaxies 
or by different missing galaxy rates inside the groups. 
To distinguish in this question, we split in each snapshot the faint galaxies in each group 
into two categories: old members and new members. We define as 'old' those 
galaxies that have already been identified as members of the same FoF group in 
the previous snapshot (higher z), while the 'new' members are those that were 
identified as members only in the corresponding snapshot. 
We also detected galaxies that existed in a previous snapshot and 
were missing in the following ('lost' members). 
In the right plot of Fig.~\ref{f4} we show the 
mean number of new and lost faint galaxy members
in fossils and controls as a function of redshift. 
In the range from $z\simeq0.7$ to $z\simeq 0.3$
fewer new faint members are incorporated in each snapshot in both 
fossils and controls. However, this decrease is stronger in fossils. 
For $z\lesssim 0.3$ there are fewer lost galaxies 
in fossils than missing galaxies in controls. 
These results remain valid for both fossil samples.
 
The almost constant number of faint galaxies observed in fossils in 
the range $0.3 < z < 0.6$ (left plot) could be a consequence of a stronger decrease of 
the accretion rate of faint galaxies in fossils at these redshifts. 
On the other hand, the rapid growth of the number of faint galaxies
in fossils, which ultimately causes the number of faint galaxies in fossils and controls 
to be the same at redshift zero, 
could be a consequence of a smaller number of missing galaxies
in fossils since $z\simeq 0.3$.

\section{Summary and conclusions}
\label{theend}
We have deepened our analysis on fossil groups from the
semi-analytical point of view with a higher resolution than in previous works, 
by studying the evolution of the main properties
of fossil systems and the distribution of the 
faint galaxy population that inhabits these peculiar systems. 

The work was based on semi-analytical galaxies constructed by \cite{guo11} based on
the high-resolution N-body numerical simulation, the Millennium run simulation II
\citep{mII}. These mock galaxies are a highly suitable tool to investigate the
role of dwarf galaxies embedded in larger density structures.
Although our study is based on only one particular
set of semi-analytical galaxies, that is, on the adopted set of recipes
defined by \cite{guo11}, this sample is one of the largest samples of faint
galaxies up to date, and their recipes have been tuned specifically to improve the
evolution of satellite galaxies compared to previous versions of the semi-analytical 
models of galaxy formation. They also carefully reproduce observational
results such as the galaxy luminosity function and the stellar mass distribution
 in a very wide range of absolute magnitudes and stellar masses. 

We identified fossil groups in our mock
catalogue and then followed their evolution back in time for the past 
$\sim 8$ Gyr ($z\sim1$).  We adopted two different definitions of fossil 
systems that can be found in the literature:
the well-known definition of \cite{jones03}, which is based on the absolute magnitude gap 
between the first and second brightest galaxies in the 
group within half a virial radius; 
and the definition of \cite{dariush10}, which is based on the absolute magnitude gap 
between the first and fourth brightest galaxies (see Sect.~\ref{fossilsec} 
for details).
Our intention in using these two criteria was to investigate whether using
different definitions affects the evolution of the brightest and
faintest galaxies in fossil systems. 
Fossil groups also met the requirement that their virial masses are higher than 
$10^{13.5} \, \rm h^{-1} \, \cal{M}_\odot $ 
and that they assembled half of their virial masses at least 7 Gyr ago. 
The latter constraint led to samples of fossil groups that represents
 $64 - 68\%$ of optical fossil groups 
(selected only according to group mass and magnitude gap). 
We have also defined reliable samples of control groups for
each criterion to perform a fair comparison.
These control groups have the same distributions of 
virial masses as fossil groups and similar early assembly times, 
 but present a much smaller gap between the magnitudes of their 
brightest galaxies. 

First, we analysed the evolution of the main properties of the brightest galaxies 
in fossil and control groups.
We observed that the brightest galaxy in fossil groups 
is typically brighter and more massive than their counterparts in control groups. 
From our studies, it is clear that fossil groups start fulfilling their
criterion to be considered as such around $z\simeq 0.2-0.3$ ($\Delta_{12}>2$ or $\Delta_{14}>2.5$). 
We note that fossil groups defined using the \cite{dariush10} criterion 
start being fossils earlier than with the usual criterion, 
and therefore they  maintain their fossil phase longer than 
with the definition of \citealt{jones03}, as stated in the work by \citealt{dariush10}.
We found that the brightest galaxy in fossils defined with the classical criterion assembled 
half of their current stellar mass at later times than the brightest galaxies in 
fossils defined with the alternative criterion or in control groups.
Finally, we observed that the stellar masses of the brightest galaxies of 
all types of groups show a notorious increment around $z\simeq 0.3$. 
In a previous work \citep{diaz08}, 
it has been shown that the brightest galaxy in fossil groups has assembled half of its
final mass later than non-fossil brightest galaxies 
and that it has experienced the latest major merger relatively recently ($z\sim 0.3$) 
compared to the brightest galaxies with similar stellar mass hosted in normal groups. 
Here, we found that the brightest galaxy in fossils and controls 
had a major event at the same redshift. 
However, it has to be noted that the brightest galaxies in fossils in this work
are more massive than their counterpart in controls, which leads to the difference 
with the previous findings. It is interesting to note that also around the time when 
the major event has happened in both group types, 
the magnitude gap in fossil groups increases 
significantly, while the magnitude gap in controls is not affected. 
This change in the magnitude gap arises from a drastic change in the magnitude 
of the second-ranked galaxy in fossils. By analysing
the behaviour of the magnitude gap between the first- and 
second-, third-, and fourth-ranked galaxies,
we could infer that at earlier times, 
fossil groups comprised two large brightest galaxies with 
similar magnitudes surrounded by much
fainter galaxies, and that at a redshift of around $0.2 - 0.3$ 
these two galaxies merged to form the brightest galaxy that we observe today. 
This event probably was a dry merger given the non-significant 
change in the absolute magnitude of the brightest galaxy. This assumption agrees with the hypothesis of \cite{fogo2}, who stated that the brightest galaxies
of fossil groups had suffered wet mergers only at early times, while the bulk of their mass 
is assembled through subsequent dry mergers. 
On the other hand, the brightest galaxies in control groups were not as massive as 
those in fossils, and the differences in magnitude with the 
second-, third-, and fourth-ranking galaxies were not as large as in fossils; 
therefore, after the merger event that they have also experienced at 
around $z\sim0.3$, the magnitude gap is not as affected as in fossils. 

The different luminosity sampling of fossil and control groups at early times 
could be a consequence of different merging histories before the period of time 
analysed in this work ($\gtrsim 8 $ Gyr). \cite{burke13} analysed the brightest 
galaxies of clusters at $z\sim1$ and demonstrated  that 
similar mass clusters have very different merging histories.
They also stated
that both major and minor mergers were more common in the past. Therefore, we suggest
that these merging scenarios must have been more efficient in fossils than in controls, 
where the two brightest central galaxies have accreted their bright companions, 
leading to more suitable initial conditions to the formation of the large magnitude 
gap at later times. Moreover, according to these arguments, 
it is not probable that control groups will eventually develop a large
magnitude gap from merging since the rate of mergers at later times is much lower. This
scenario reinforces the idea that fossil groups are a different type of systems.

Second, we studied the faint galaxy population ($-16 \le M_r - 5 \log{(\rm h)} \le -11$) 
in fossil and control groups by analysing their number density profiles around 
the group centres. We observed that at $z=0$, fossil systems show a very similar galaxy
density profile of faint galaxies when compared with their corresponding control samples.
Nevertheless, when analysing their evolution with time, some differences between 
faint galaxies in fossil and control groups stood out:  
at earlier times ($z\gtrsim 0.7$), the population of faint galaxies
in fossil systems is denser than observed in control groups in 
a wide range of distances to the centre. At later times ($0.3\lesssim z\lesssim 0.5$), 
the previous trends are reversed, and control groups appears to be denser than 
fossil systems.
In the same period of time, 
the mean number of faint galaxies in fossil groups remains roughly constant, 
while in control groups it continues growing. 
This almost constant number of faint galaxies observed in fossils
could be a consequence of a strong decrease of 
the accretion rate of faint galaxies at these redshifts. 
For $z\lesssim 0.2$, the mean number of faint galaxies in fossil systems grows
to finally reach the values observed in control groups at $z=0$.
This later result agrees with the observational work of \cite{lieder13},
who found a normal abundance of faint satellites in the NGC 6482 fossil group.
This rapid growth in the number of faint galaxies
could be a consequence of a smaller number of missing galaxies in fossils. 

\cite{goz14} analysed the number of galaxies in the range of 
magnitudes between $-18$ to $-16$ that inhabit fossil and normal groups 
in the MS-I simulation and found no evolution in this population 
in fossil groups,
 while they observed an increase of $\sim 40\%$  in normal groups towards low redshifts. 
In this work, we extended their analysis to fainter galaxies
and found that the number of faint galaxies grows in both fossils and controls
since $z\sim1$ to the present day, although this growth occurred in 
different ways, as we explained above.

We conclude that using either of the two different criteria to define fossil systems
does not have a very strong effect on the evolution of the brightest or 
faintest populations. We confirm that the definition of \cite{dariush10} 
allows detecting systems that became fossils earlier, and 
the assembly time of its brightest galaxy occurred before the brightest galaxy
that inhabits fossil groups selected with the classical definition of \cite{jones03}.

The predictions presented in this work need to be confirmed 
when more observational data including fainter galaxies
 are available, and/or  when larger high-resolution semi-analytical galaxy samples 
for different models of galaxy formation are released. 

\begin{acknowledgements}
The Millennium Simulation databases used in this paper and the web application 
providing online access to them were constructed as part of the activities of 
the German Astrophysical Virtual Observatory (GAVO). We thank Qi Guo for allowing 
public access for the outputs of her very impressive semi-analytical model of 
galaxy formation. This work has been partially supported by Consejo Nacional 
de Investigaciones Cient\'\i ficas y T\'ecnicas de la Rep\'ublica Argentina 
(CONICET) and the Secretar\'\i a de Ciencia y Tecnolog\'\i a de la Universidad 
de C\'ordoba (SeCyT).
\end{acknowledgements}

%%%%%%%%%%%%%%%%%%%%%%%%%%%%%%%%%%%%%%%%%%%%%%%%%%%%%%%%%%%%%%%%%%%%%%%%%%%%%%%
\bibliography{refs}
%%%%%%%%%%%%%%%%%%%%%%%%%%%%%%%%%%%%%%%%%%%%%%%%%%%%%%%%%%%%%%%%%%%%%%%%%%%%%%%

\appendix

\end{document}